\documentclass[twocolumn,aps,superscriptaddress,showpacs,floatfix]{revtex4}
\usepackage{graphicx}
\usepackage{dcolumn}
\usepackage{bm}
\usepackage[colorlinks,linkcolor=blue,anchorcolor=blue,citecolor=blue,urlcolor=blue]{hyperref}
\usepackage{amsmath}
\usepackage{multirow}
\usepackage{booktabs}  
\usepackage{longtable}  
\usepackage{float}
\usepackage{array}
\usepackage{etoolbox}
\usepackage{makecell,rotating,multirow,diagbox}
\usepackage{setspace}
\usepackage{tabularx}
\usepackage[draft]{changes}
\usepackage[section]{placeins}
\usepackage{lineno,hyperref}
\usepackage{supertabular}
\usepackage{amssymb}
\usepackage[normalem]{ulem}
\usepackage[colorlinks,linkcolor=blue,anchorcolor=blue,citecolor=blue,urlcolor=blue]{hyperref}
\renewcommand\sout{\bgroup \color{red} \ULdepth=-.5ex \ULset}

\begin{document}

\title{Cluster radioactivity preformation probability of trans-lead nuclei in the scheme of N$_{p}$N$_{n}$}

\author{Lin-Jing Qi}
\affiliation{School of Nuclear Science and Technology, University of South China, 421001 Hengyang, People's Republic of China}
\author{Dong-Meng Zhang}
\affiliation{School of Nuclear Science and Technology, University of South China, 421001 Hengyang, People's Republic of China}
\author{Song Luo}
\affiliation{School of Nuclear Science and Technology, University of South China, 421001 Hengyang, People's Republic of China}
\author{Gui-Qing Zhang}
\email{nkzhanggq@tust.edu.cn}
\affiliation{College of Science, University of Science and Technology, 300457 Tianjin, People's Republic of China}
\author{Peng-Cheng Chu}
\email{kyois@126.com}
\affiliation{The Research Center for Theoretical Physics, Science School, Qingdao University of Technology, Qingdao 266033, China}
\author{Xi-Jun Wu}
\email{wuxijunusc@163.com}
\affiliation{School of Math and Physics, University of South China, Hengyang 421001, People's Republic of China}
\author{Xiao-Hua Li}
\email{lixiaohuaphysics@126.com }
\affiliation{School of Nuclear Science and Technology, University of South China, 421001 Hengyang, People's Republic of China}
\affiliation{Cooperative Innovation Center for Nuclear Fuel Cycle Technology $\&$ Equipment, University of South China, 421001 Hengyang, People's Republic of China}
\affiliation{National Exemplary Base for International Sci $\&$ Tech. Collaboration of Nuclear Energy and Nuclear Safety, University of South China, Hengyang 421001, People's Republic of China}

\begin{abstract}
	
In the present work, the cluster radioactivity preformation probability $P_{c}$ in the scheme of $N_{p}N_{n}$ for the effective number of the valence particles (holes) in trans-lead nuclei has been systematically investigated. This quantity has been explored in the simplified parametrization of $N_{p}N_{n}$ as well as the multiplication $N_{p}N_{n}I$ of this product with the isospin asymmetry $I$. The calculations for $P_{c}$ are both performed in microscopic and model-dependent way. Within the microscopic approach, based on our previous work [\href{https://doi.org/10.1088/1674-1137/ac94bd}{Chin. Phys. C \textbf{47}, 014101 (2023)}], $P_{c}$ is calculated in cluster formation model (CFM) combined with the exponential relationship of $P_{c}$ to the $\alpha$ decay preformation probability $P_{\alpha}$ when the mass number of the emitted cluster $A_{c}$ $\leq$ 28. While $A_{c}$ $\ge$ 28, $P_{c}$ is obtained through the charge-number dependence of $P_{c}$ on the decay products proposed by Ren $et$ $al.$ [\href{https://doi.org/10.1103/PhysRevC.70.034304}{Phys. Rev. C  \textbf{70}, 034304 (2004)}]. In the model-dependent approach, $P_{c}$ is extracted through the ratios from calculated cluster radioactivity half-lives in the \replaced{framework of }{} unified fission model (UFM) proposed by Dong $et$ $al.$ [\href{https://doi.org/10.1140/epja/i2009-10819-1}{Eur. Phys. J. A \textbf{41}, 197 (2009)}] to experimental ones. Both of the results show $P_{c}$ in logarithmic form are linear to $N_{p}N_{n}$ as well as $N_{p}N_{n}I$. For comparison, the parent-mass-number dependence analytical formula as well as the model proposed by K. Wei and H. F. Zhang [\href{https://doi.org/10.1103/PhysRevC.96.021601}{Phys. Rev. C  \textbf{96}, 021601(R) (2017)}] are also used. Furthermore, the preformation mechanic for cluster radioactivity has also been discussed.

\end{abstract}

\pacs{21.60.Gx, 23.60.+e, 21.10.Tg}
\maketitle

\section{Introduction}
\setlength{\parskip}{0pt}

 Nuclear physics was originated from the discovery of natural radioactivity.
 Nuclear spontaneous disintegration has always been the effective probe for investigating nuclear structure. In 1980, S$\check{a}$ndulescu, Poenaru and Greiner primarily predicted a novel spontaneous emission phenomenon in unstable nuclei whose emitted fragments, heavier than $\alpha$ particles but less than fission fragments, generally are known as cluster radioactivity \cite{J.Phys.G:Nucl.Part.Phys.15 529(1989),Int. J. Mod. Phys. E 3 335 (1994),J. Phys. G: Nucl. Part. Phys. 17 S443,J. Phys. G: Nucl. Part. Phys. 30 945-955,Sov. J. Part. Nucl. 11 528 (1980)}. Soon afterwards, this decay mode was experimentally confirmed by Rose and Jones through observing $^{14}$C particle emitted from $^{223}$R$a$ isotope \cite{Nature 307 245 (1984),Phys. Scr. 86 015201,Ann. Phys. 334 280 (2013),Phys.Rev.C 70 017301}. Since then, an increasing number of clusters heavier than $^{14}$C particle such as $^{20}$O, $^{23}$F, $^{22, 24-26}$Ne, $^{28, 30}$Mg and $^{32, 34}$Si isotopes are availably observed on experiments in the parent nuclei ranging from $^{221}$Fr up to $^{242}$Cm in trans-lead region decaying to the doubly magic nucleus $^{208}$Pb or its neighbouring nuclei \cite{Phys. Rev. C 97 014318,Phys. Rev. C 34 2261,Eur. Phys. J. A 49 6,Phys.Rev.C 83 014601}, further providing a unique way to explore various nuclear structures. 
 
 Since cluster radioactivity, proton radioactivity, two-proton radioactivity and $\alpha$ decay are similarly explained as the quantum mechanical effect \cite{Eur. Phys. J. A 55 214,Phys.Rev.C 80 044326,Nucl. Phys. A 958 202,Chin. Rhys. C 46 044106,Eur. Rhys. J. A 58 244,Nucl. Sci. Tech. 34 55,Nucl. Sci. Tech. 33 122,Nucl. Sci. Tech. 34 30,Eur. Rhys. J. A 58 16,Phys. Scr. 96 075301,Chin. Rhys. C 44 094106}, in particular, due to the immediate characteristic between $\alpha$ decay and spontaneous fission, there are generally two kinds of theoretical methods within Gamow's theory well established to interpret this rare decay mode: $\alpha$-like models and fission-like models \cite{J.Phys.G 39 095103}. Furthermore, the cluster radioactivity preformation probability is dealt with differently in these two models \cite{Int. J. Mod. Phys. E 23 1450018,Phys. Rev. C 85 054612 (2012),Nucl. Phys. A 992 121626 (2019),Sov. J. Part. Nucl. 11 528 (1980)}. In $\alpha$-like models, the cluster is assumed to already be pre-born in the parent nuclei with certain probability before penetrating the interacting barrier between the emitted cluster and the daughter nucleus \cite{Phys. Scr. 96 125322,Chin. Phys. C 46 044104,Phys. Rev. C 71 024308,Nucl. Phys. A 683 182 (2001)}. For actually, in density-dependent cluster model (DDCM) proposed by Ren \emph{et al.}, the cluster preformation probability is assumed as an exponential function as the multiplication of the emitted cluster charge number as well as daughter charge number employed to calculate the half-lives of cluster radioactivity \cite{Phys.Rev.C 70 034304,Phys. Rev. C 78 044310}. In preformed cluster model (PCM), it is obtained by solving the stationary Schr$\ddot{o}$dinger equation for the dynamical flow of mass and charge \cite{Phys. Rev. C 86 044612}. However, in fission-like models, cluster radioactivity preformation probability is regarded as the penetration probability of the pre-scission part for the interacting barrier \cite{Phys. Rev. C 32 572,Chin. Phys. C 45 044111}. For instance, in coulomb and proximity potential (CPPM)
proposed by Santhosh \emph{et al.} \cite{J. Phys. G: Nucl. Part. Phys. 35 085102}, the cluster formation probability is calculated as the penetrability probability through the internal part of the potential barrier through the simple power-law interpolation. 

Noticeably, $P_{c}$ plays an indispensable incorporated part in calculating cluster radioactivity half-lives \cite{Int. J. Mod. Phys. E 31 2250068}. Moreover, it is an important indicator possessing abundant information of nuclear structure such as shell effects, surface deformation and neutron--proton (n-p) interaction in our previous study and investigation from other researchers \cite{Chin. Phys. C 47 014101,Phys. Scr. 95 075303,Phys. Rev. C 39 1992,Phys. Rev. C 96 021601}. Since cluster radioactivity is closely related to shell effects and $P_{c}$ can be considered as the penetrability for the overlapping region between the actual ground state configuration of the parent and the configuration described by the emitted cluster coupled to the ground state of the daughter nucleus \cite{Phys. Rev. C 102 034318}, the tunneling probability of the emitted cluster is expected to be significantly dependent on the n-p interaction. Various quantities such as deformation, ground band energy systematics and properties of excited states can be parameterized in the scheme of $N_{p}N_{n}$ or obtained through the simple functions of $N_{p}$ and/or $N_{n}$ which may also bear smooth relationships with the observables \cite{Phys. Lett. B 665 182}. n-p interaction can be well represented by these parameterization and simple functions \cite{Phys. Lett. B 665 182,J. Phys. G: Nucl. Part. Phys. 36 095105}. Previous work has indicated the general behavior of $P_{c}$ in logarithmic form linearly decrease with the neutron number arriving at a local minimum as the magic shell closures and then linearly increase again \cite{Phys. Scr. 95 075303}. This phenomenon may imply $P_{c}$  exhibits a certain correlation when it is expressed as a function of the product of $N_{p}$ and $N_{n}$. In the present work, we make attempts to explore the relationship of cluster radioactivity preformation probability versus valence nucleons (holes) in the scheme of $N_{p}$$N_{n}$, while cluster radioactivity preformation probability is calculated both microscopically and model-dependently. In the microscopic approach, based on our previous work, the cluster radioactivity preformation probability is dealt with CFM combined with the exponential relationship of $P_{c}$ to the $\alpha$ decay preformation probability $P_{\alpha}$ when the number of the emitted cluster $A_{c}$ $\leq$ 28. It should be noted that, as is clearly indicated in Fig. 2 from Ref \cite{Phys. Rev. C 96 021601}, $P_{c}$ in logarithmical form keeps a good linear relationship with the mass number of the emitted cluster. The curve is bent obviously when $A_{c}$ $>$ 28 and the slope of the curve begin to decrease with the increasing of the emitted cluster mass number. Therefore, in the present work, while $A_{c}$ $\ge$ 28, $P_{c}$ is obtained through the charge-number dependence of $P_{c}$ on the decay products proposed by Ren $et$ $al.$ \cite{Phys.Rev.C 70 034304}. In the model-dependent approach, $P_{c}$ is extracted through the ratios from the calculated cluster radioactivity half-lives to experimental ones, while the cluster radioactivity half-lives calculations are preformed in the unified fission model (UFM) \cite{Eur. Phys. J. A 41 197}. 

This article is organized as follows. A brief introduction of the theoretical framework for UFM and CFM is briefly presented in Section \ref{section 2}. Detailed numerical results and discussion are given in Section \ref{section 3}. Section \ref{section 4} is a simple summary. 

\section{Theoretical framework}
\label{section 2}
\subsection{Model-dependent approach}
In UFM, for the emitted cluster-daughter system, the barrier penetration probability $P$ can be obtained by the Wentzel-Kramers-Brillouin (WKB) approximation action integral \cite{Eur. Phys. J. A 41 197}
\begin{equation}
P={exp}{\lbrace -\frac{2}{\hbar}\int_{R_{in}}^{R_{out}}\sqrt {2\mu (V(r)-Q_{c}) }\mathrm{d}r\rbrace}\replaced{,}{.}
\end{equation}
where $\hbar$ is the reduced Planck constant and $\mu$ =$\frac{M_{c} M_{d}}{M_{c}+M_{d}}$ is the reduced mass of emitted cluster-daughter nucleus system with $M_{c}$ and $M_{d}$ being the masses of emitted cluster and daughter nucleus, respectively \cite{Nucl. Phys. A 951 86}. $Q_{c}$ is the cluster radioactivity decay energy.  It can be obtained by \cite{Mod. Phys. Lett. A 30 1550150 (2015)}
\begin{equation}
Q_{c}=B(A_{c},Z_{c})+B(A_{d},Z_{d})-B(A,Z),   
\end{equation}
where $B(A_{c},Z_{c})$, $B(A_{d},Z_{d})$ and $B(A,Z)$ are, respectively, the binding energy of the emitted cluster, daughter and parent nuclei taken from AME2020 \cite{Chin. Phys. C 45 030001} and NUBASE2020 \cite{Chin. Phys. C 45 030003}. $A_{c}$, $Z_{c}$, $A_{d}$, $Z_{d}$ and $A$, $Z$ are the mass numbers and proton numbers of the emitted cluster, daughter and parent nucleus, respectively. 
$R_{in}$ = $R_{1}$ + $R_{2}$ and $R_{out}=\frac{Z_{c}Z_{d}e^{2}}{2Q_{c}}+\sqrt{(\frac{Z_{c}Z_{d}e^{2}}{2Q_{c}})^{2}+\frac{l(l+1)\hbar^{2}}{2\mu Q_{c}}}$ are the radius for the separation configuration and the outer turning point \cite{Eur. Phys. J. A 41 197} with $R_{1}$ and $R_{2}$ being the equivalent sharp radii of the daughter nucleus and the emitted cluster, respectively. They can be obtained by \cite{Mod. Phys. Lett. A 30 1550150 (2015)}
\begin{equation}
R_{i}=(1.28A_{i}^{1/3}-0.76+0.8A_{i}^{-1/3})fm, i=1,2.
\end{equation}

The total interacting potential $V(r)$ between the emitted cluster and  daughter nucleus is consisted of the repulsive long-range Coulomb potential $V_{C}(r)$, the attractive short-range nuclear proximity potential $V_{p}$ and the centrifugal potential $V_{l}(r)$ when the fragments are separated. It is expressed as \cite{Eur. Phys. J. A 41 197}
\begin{equation}\label{eq4}
V(r)=V_{p}(r)+V_{C}(r)+V_{l}(r),
\end{equation}
where $r$ is the distance between the fragment centers. The inclusion of the proximity potential reduces the height of the barrier which closely agrees with the experimental values. The nuclear proximity potential takes the following form \cite{Eur. Phys. J. A 41 197},
\begin{equation}
V_{p}(r)=4\pi\frac{C_{1}C_{2}}{C_{1}+C_{2}}\gamma b\Phi(s),
\end{equation}
where the S$\ddot{u}$smann central radii  $C_{i}$ = $R_{i}$-$\frac{b^{2}}{R_{i}}$ with $b$=0.99 fm being the surface width. The nuclear surface tension coefficient $\gamma$ is given as \cite{Phys. Rev. C 85 054612 (2012)}
\begin{equation}
\gamma=0.9517[1-1.7826(\frac{N-Z}{A})^{2}]MeV \cdot fm^{-2}, 
\end{equation}
where $N$ represents the neutron number of the parent nucleus. The universal function $\Phi(s)$ is parameterized as \cite{Phys. Rev. C 85 054612 (2012)}
\begin{eqnarray}
\Phi(s)=\left\{\begin{array}{llll}
\frac{1}{2}(s-2.54)^{2}-0.00852(s-2.54)^{3}, &\rm{s \leq 1.2511}, \\

-3.437\exp(-\frac{s}{0.75}), &\rm{s \ge 1.2511},
\end{array}\right.
\end{eqnarray}
where $s=(r-C_{1}-C_{2})/b$ is the overlap distance in units
of $b$ for the colliding surfaces. 

The Coulomb potential in Eq.\ref{eq4} is given as \cite{Chin. Phys. C 47 014101}
\begin{equation}\label{8}
V_{C}(r)=\frac{e^{2}Z_{c}Z_{d}}{r},
\end{equation}
where  $e^2=1.4399652 MeV\cdot fm$ is the square of the  electronic elementary charge \cite{Chin. Phys. C 47 014101}. As for the centrifugal potential $V_{l}(r)$, since $l(l+1)$ $\to$ $(l+\frac{1}{2})^2$ is a necessary correction for one-dimensional problems \cite{Phys. Rev. C 82 024311}, we choose it as the Langer modified form in this work. It can be expressed as
\begin{equation}
V_{l}(r)=\frac{\hbar^{2}(l+\frac{1}{2})^{2}}{2\mu r^{2}},
\end{equation}
where $l$ is the angular momentum carried by the emitted cluster. 
It can be obtained by \cite{Phys. Rev. C 97 044322}
\begin{eqnarray}\label{14}
l=\left\{\begin{array}{llll}
\Delta_{j}, &\rm{for \ even \ \Delta_{j} \ and \ \pi_{p}=\pi_{d}}, \\
\Delta_{j}+1, &\rm{for \ even \ \Delta_{j} \ and \ \pi_{p}\neq\pi_{d}},\\
\Delta_{j}, &\rm{for \ odd\ \Delta_{j} \ and \ \pi_{p}\neq\pi_{d}}, \\
\Delta_{j}+1, &\rm{for \ odd \ \Delta_{j} \ and \ \pi_{p}=\pi_{d} },
\end{array}\right.
\end{eqnarray}
where $\Delta_{j}=\lvert j_{p}-j_{d}-j_{c} \rvert$,  $j_{c},\pi_{c},j_{p},\pi_{p}$ and $j_{d},\pi_{d}$ are the isospin and parity values of the emitted cluster, parent and daughter nuclei, respectively.

The assault frequency $\nu_{0}$ is calculated by \cite{Eur. Phys. J. A 41 197}
\begin{equation}
\nu_{0}=\frac{1}{R_{0}}\sqrt{\frac{2E}{M}},
\end{equation}
where $E$ and $M$ are the kinetic energy and  mass of the emitted cluster, respectively. With the experimental cluster radioactivity half-life, the preformation probability can be extracted from \cite{Eur. Phys. J. A 41 197}
\begin{equation}\label{Eq.9}
P_{c}=\frac{\ln 2}{T_{exp}\nu_{0}P}.
\end{equation}

\subsection{Microscopic approach}

Within the framework of CFM, the total initial clusterization state $\psi$ of the emitted cluster-daughter nucleus system is considered as a linear superposition of all its $n$ possible clusterization $\psi_{i}$ states \cite{J. Phys. G: Nucl. Part. Phys. 40 065105}. It can be expressed as
\begin{equation}
\psi=\sum_{i}^{N}a_{i}\psi_{i},
\end{equation}
\begin{equation}
a_{i}=\int{\psi_{i}}^{*}\psi\mathrm{d}\tau,
\end{equation}
 where $a_{i}$ represents the superposition coefficient of $\psi_{i}$ which satisfies the orthogonality condition \cite{Nucl. Phys. A 962 103 (2017)}
\begin{equation}
\sum_{i}^{N}\lvert a_{i} \rvert^{2}=1.
\end{equation}
Correspondingly, the total Hameiltonian $H$ is consisted of the Hameiltonian $H_{i}$ for different clusterization configuration $\psi_{i}$ which can be expressed as \cite{J. Phys. G: Nucl. Part. Phys. 42 075106}
\begin{equation}
H=\sum_{i}^{N}H_{i}.
\end{equation} 

By virtue of all the clusterization states describing the same emitted cluster-daughter nucleus system, they are assumed as sharing the same total eigen-energy $E$ of the total wave function \cite{Phys. Rev. C 93 044326}. Furthermore, considering the orthogonality of the clusterization wave functions, $E$ can be expressed as 
\begin{equation}
E=\sum_{i}^{N}\lvert a_{i} \rvert^{2}E=\sum_{i}^{N}E_{f_{i}},
\end{equation}
 where $E_{f_{i}}$ represents the formation energy for the cluster in the $i$-th clusterization state $\psi_{i}$. For $\alpha$ decay, the preformation probability $P_{\alpha}$ can be obtained by \cite{J. Phys. G: Nucl. Part. Phys. 42 075106}
\begin{equation}
P_{\alpha}=\lvert a_{\alpha} \rvert^{2}=\frac{E_{f_{\alpha}}}{E}.
\end{equation}
Here $a_{\alpha}$ and $E_{f_{\alpha}}$ are the coefficient of the $\alpha$ clusterization state and the formation energy of the $\alpha$ particle, respectively. 

Moreover, the $\alpha$ formation energy $E_{f_{\alpha}}$ and total system energy $E$ can be classified as four different cases in the following expressions \cite{Phys. Rev. C 97 044322}.

Case
$\uppercase\expandafter{\romannumeral1}$ for even-even nuclei
\begin{equation}
\begin{split}
E_{f_{\alpha}}=3B(A,Z)+B(A-4,Z-2)\\
-2B(A-1,Z-1)-2B(A-1,Z),
\end{split}
\end{equation}
\begin{equation}
E=B(A,Z)-B(A-4,Z-2).
\end{equation}

Case $\uppercase\expandafter{\romannumeral2}$ for even-odd nuclei
\begin{equation}
\begin{split}
E_{f_{\alpha}}=3B(A-1,Z)+B(A-5,Z-2)\\
-2B(A-2,Z-1)-2B(A-2,Z),
\end{split}
\end{equation}
\begin{equation}
E=B(A,Z)-B(A-5,Z-2).
\end{equation}

Case $\uppercase\expandafter{\romannumeral3}$ for odd-even nuclei:
\begin{equation}
\begin{split}
E_{f_{\alpha}}=3B(A-1,Z-1)+B(A-5,Z-3)\\
-2B(A-2,Z-2)-2B(A-2,Z-1),
\end{split}
\end{equation}
\begin{equation}
E=B(A,Z)-B(A-5,Z-3).
\end{equation}

Case $\uppercase\expandafter{\romannumeral4}$ for odd-odd nuclei:
\begin{equation}
\begin{split}
E_{f_{\alpha}}=3B(A-2,Z-1)+B(A-6,Z-3)\\
-2B(A-3,Z-2)-2B(A-3,Z-1),
\end{split}
\end{equation}
\begin{equation}
E=B(A,Z)-B(A-6,Z-3).
\end{equation}
When A$_{c}$ $<$ 28, $P_{c}$ in logarithmic form keeps a good linear relationship with $P_{\alpha}$. Using this relation, $P_{c}$ can be obtained by \cite{Phys. Rev. Lett. 61 1930 (1988)}
\begin{equation}
P_{c}=[P_{\alpha}]^\frac{(A_{c}-1)}{3}.
\end{equation}
As for $A_{c}$ $>$ 28, the calculations for the $P_{c}$ are completed by a  formula proposed by Ren $et.al$ \cite{Phys.Rev.C 70 034304} for this law may not work. It can be expressed as
\begin{eqnarray}\label{eq 17}
log_{10} P_{c}=\left\{\begin{array}{llll}
-(0.01674Z_{c}Z_{d}-2.035466),\\
\rm{for \ even-even \ nuclei}\\
-(0.01674Z_{c}Z_{d}-2.035466)-1.175,\\
\rm{for \ odd-A \ nuclei}.
\end{array}\right.
\end{eqnarray}
\section{Results and discussion}
\label{section 3}
\begin{figure}[h]\centering
	\includegraphics[width=8.6cm]{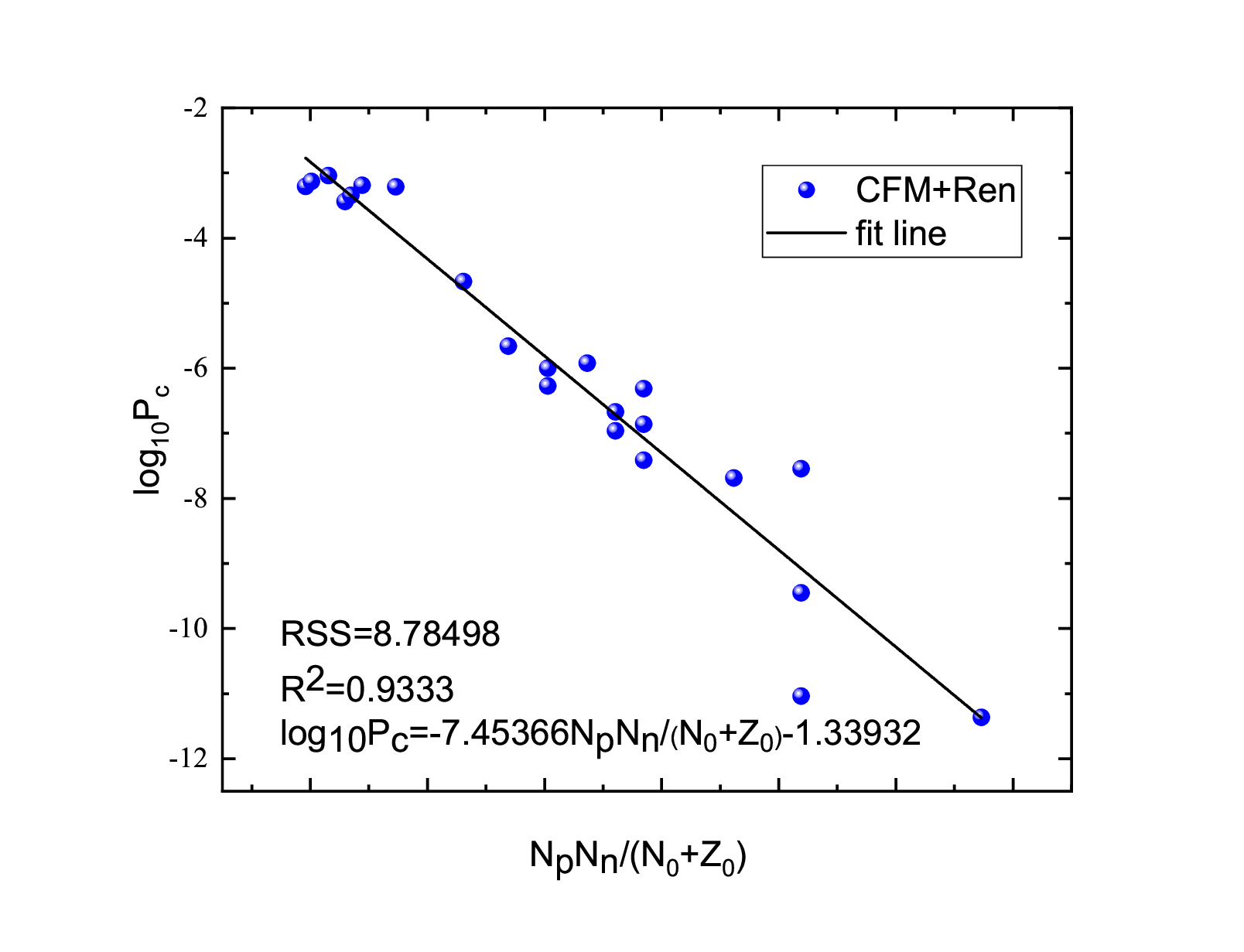}
	\caption{(color online) The linear relationship of the cluster radioactivity preformation probability in logarithmic form by using microscopic approach with the product of the valance protons (holes) and neutrons (holes) $NpNn$.}
	\label{fig 1}
\end{figure}

\begin{figure}[h]\centering
	\includegraphics[width=8.6cm]{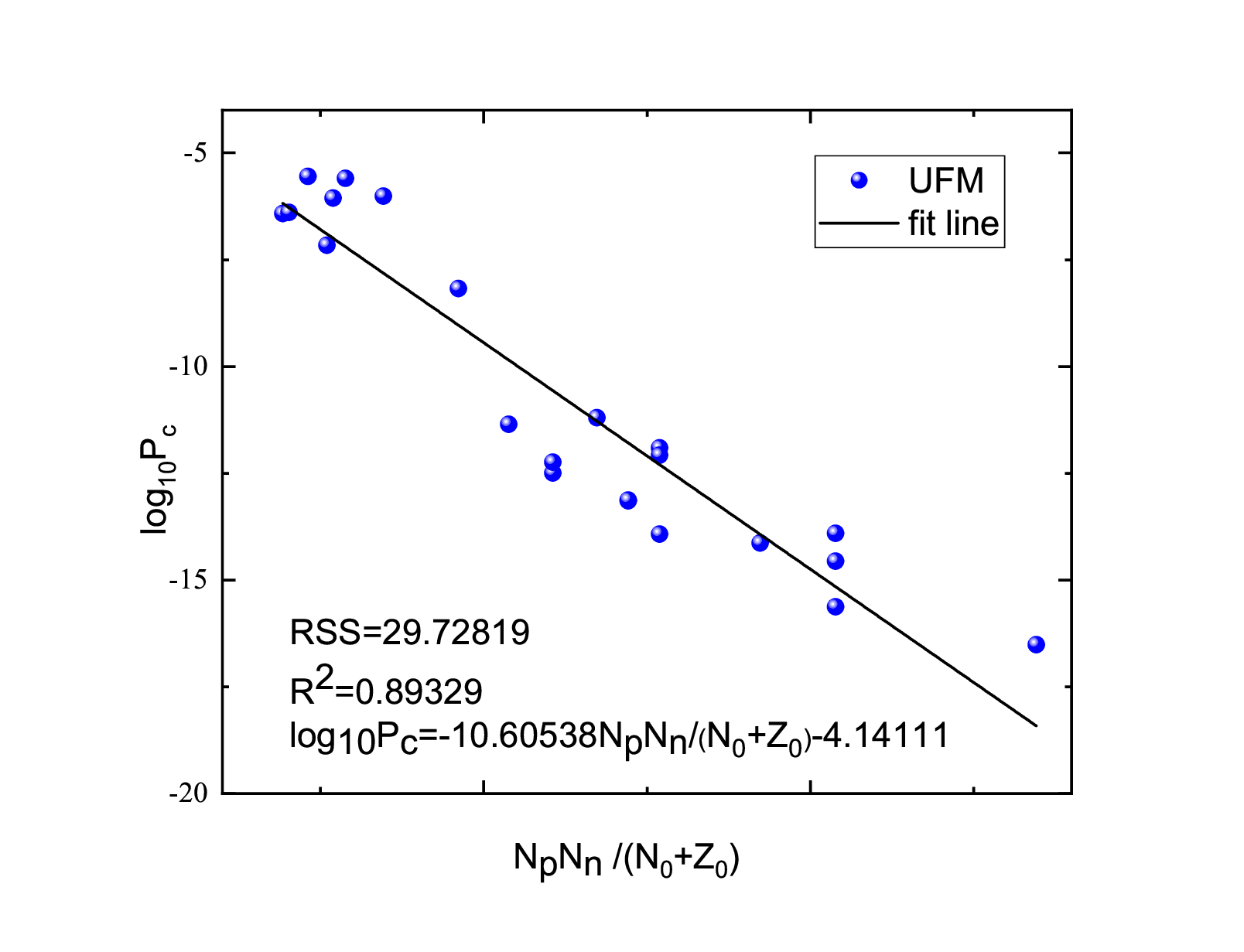}
	\caption{(color online) The linear relationship of the cluster radioactivity preformation probability in logarithmic form  by using model-dependent approach with the product of the valance protons (holes) and neutrons (holes) $NpNn$.}
	\label{fig 2}
\end{figure}

The aim of this work is to systematically investigate the behavior of cluster radioactivity preformation probability of trans-lead nuclei in the scheme of $N_{p}N_{n}$. Numerous researchers have discovered such dependence in $\alpha$ decay. For instance, the works of Seif \emph{et al.} reported that $\alpha$ decay preformation probability of even-even nuclei around the $Z = 82, N = 126$ closed shells linearly depend on the product of the valance protons(holes) and neutrons (holes) $NpNn$ \cite{Phys. Rev. C 84 064608,J. Phys. G: Nucl. Part. Phys. 40 105102}. Furthermore, in our previous works, we systematically studied the $P_{\alpha}$ of the favored and unfavored $\alpha$ decay for odd-A and doubly odd nuclei, where $P_{\alpha}$ is extracted from the ratios of calculated $\alpha$ decay half-lives to experimental values \cite{Phys. Rev. C 94 024338,Chin. Rhys. C 44 094106}. The results indicated that $P_{\alpha}$ is linearly to $NpNn$ although it is model dependent. Before long, Deng \emph{et al.} further pointed out that this linear relationship simultaneously satisfies all types of nuclei well in $\alpha$ decay \cite{Phys. Rev. C 96 024318}. For $\alpha$ decay and cluster radioactivity share the same physical mechanism, it is interesting to explore whether it is a possibility for the cluster radioactivity preformation probability to have certain correlation with the product of valance protons (holes) and neutrons (holes) $NpNn$. 
For further verifying this assumption, in the present work, we use two different approaches to deal with preformation probability of cluster radioactivity. In UFM, $P_{c}$ is usually treated as unity. Theoretically, it can be extracted by the ratios from the calculated cluster radioactivity half-lives to experimental ones. Using this method, the cluster radioactivity preformation probability is deduced model-dependently. To this end, based on our previous work, when the mass number of the emitted cluster $A_{c}$ $\leq$ 28, we calculate the preformation penetrability of cluster radioactivity through the famous exponential law of $P_{c}$ to the $\alpha$ decay preformation probability $P_{\alpha}$, while $P_{\alpha}$ is obtained within microcosmic model CFM. Whereas $A_{c}$ $\geq$ 28, the preformation factor can be obtained through the charge-number dependence of $P_{c}$ on the decay products proposed by Ren $et$ $al$. \cite{Phys.Rev.C 70 034304}. A detailed discussion about the later approach has been given in Ref. \cite{Chin. Phys. C 47 014101}. Both the calculated results are well listed in Table. \ref{Tab1}. In this table, the first to third columns represent the decay process, cluster radioactivity decay energies and experimental cluster radioactivity half-lives in logarithmic form taken from Ref. \cite{Phys. Rev. C 82 024311,J. Phys. G: Nucl. Part. Phys. 35 085102} denoted as Decay, $Q_{c}$ and $T^{exp}_{1/2}$, respectively. The effective number of valence protons and neutrons for the parent nucleus expressed as $N_{p} = Z - Z_{0}$ and $N_{n} = N - N_{0}$ are presented in fourth and fifth columns denoted as $N_{p}$ and $N_{n}$ with $Z_{0}$ and $N_{0}$ being the nearest proton and neutron closed shells, respectively. In this work, we choose ($Z_{0}$=82, $N_{0}$=126) as the considered doubly magic core for the cluster radioactivity whose decaying daughter nucleus are around the doubly magic nucleus $^{208}$Pb or its neighboring nuclei in trans-lead region. Then $N_{p}$ and $N_{n}$ can be obtained by $N_{p} = Z - 82$ and $N_{n} = N - 126$. The sixth column represents the isospin asymmetry of the parent nuclei $I = (N-Z)/(N+Z)$. The seventh and eighth columns represent the cluster radioactivity preformation factor deduced from microscopic and model-dependent approaches in logarithmic form denoted as CFM and UFM, respectively. From this table, it is obviously to see that the cluster radioactivity preformation probability $P_{c}$ values even for the order of magnitude are comparably different obtained by using above two methods while the tendency of the individual variations for $P_{c}$ values are basically consistent. The results have indicated that exploring the cluster radioactivity preformation probability in both model-dependent and microscopic way is of necessity. It is well acknowledged that conventional counting of valence protons and neutrons can be inadequate on account of change in magic number and shell structure in diverse mass regions \cite{Phys. Lett. B 665 182}. Making use of effective number of valence particles in the $N_{p}N_{n}$ scheme may significantly improve the predictability of the scheme as well as point to the emergence of new shell structure in various mass regions. In that regard, in order to have an  intuitive insight to the dependence for $P_{c}$ along with valence particles (holes), the correlation between $P_{c}$ in logarithmic form obtained through UFM as well as CFM with valence particles (holes) in the form of $\dfrac{N_{p}N_{n}}{N_{0}+Z_{0}}$ are plotted in Fig.\ref{fig 1} and Fig.\ref{fig 2}. As can be clearly seen from these two pictures, $P_{c}$ in logarithmic form are similarly varying smoothly and have linear relationships with the products of the number of the valence protons and neutrons. It can be expressed as

\begingroup
\renewcommand*{\arraystretch}{1.3}
\setlength{\tabcolsep}{0.00005mm}
\begin{longtable*}{cccccccc}	
	\caption{Cluster radioactivity preformation probability in microscopic and model-dependent approaches. See text for details.}
	\label{Tab1} \\
	\hline 
	\hline 
	{Decay}&\qquad$Q_{c}$(MeV)\qquad&\qquad\rm$T^{exp}_{1/2}$\qquad&\qquad\rm$N_{p}$\qquad&\qquad\rm$N_{n}$\qquad&\qquad $I$\qquad&\qquad CFM\qquad&\qquad UFM\qquad\\
	\hline
	\endfirsthead
	\multicolumn{8}{c}%
	{{\tablename\ \thetable{} -- continued from previous page}} \\
	\hline
	\hline 
   	{Decay}&\qquad$Q_{c}$(MeV)\qquad&\qquad\rm$T^{exp}_{1/2}$\qquad&\qquad\rm$N_{p}$\qquad&\qquad\rm$N_{n}$\qquad&\qquad $I$\qquad&\qquad CFM\qquad&\qquad UFM\qquad\\
	\hline
	\endhead
	\hline \multicolumn{8}{r}{{Continued on next page}} \\
	\endfoot
	\hline \hline
	\endlastfoot
	$^{221}$Fr$\to$ $^{207}$Tl+$^{14}$C		&	31.29 	&	14.56 	&	5	&	8	&	0.213 
	 	&	--3.204
	 		&--6.420
	 		
	\\
	$^{221}$Ra$\to$$^{207}$Pb+$^{14}$C	&	32.40 	&	$	13.39	$ 	&	$	6	$	&	7	&	0.204 	&		--3.126		&	--6.388		\\
	$^{222}$Ra$\to$$^{208}$Pb+$^{14}$C 	&	33.05 	&	11.22 &		6 	&	8 	&	0.207 	&	--3.037 	&	--5.547
	\\
	$^{223}$Ra$\to$$^{209}$Pb+$^{14}$C &	31.83 	&	15.05		&	6 	&	9 	&	0.211 	&	--3.440 	&	--7.160
	\\
	
	$^{224}$Ra$\to$$^{210}$Pb+$^{14}$C&	30.53 &	15.87	&	6 	&	10 	&	0.214 	&	--3.186 	&	--5.593 	 \\
	$^{226}$Ra$\to$ $^{212}$Pb+$^{14}$C &	28.20 	&	21.2 	&	6 	&	12 	&	0.221	&	--3.211 	&	--6.014 \\
	$^{223}$Ac$\to$$^{209}$Bi+$^{14}$C 	&	33.06 	&	12.6 	&	8 	&	12 	&	0.202	&	--3.341 	&	--6.060 \\
	$^{228}$Th$\to$$^{208}$Pb+$^{20}$O	&	44.72 	&	20.73 	&	8 	&   12 	&	0.211 	&	--4.670 	&	--8.174 \\
	$^{231}$Pa$\to$$^{208}$Pb+$^{23}$F	&	51.88 	&	26.02	& 	9 	&	14 	&	0.212 	&	--5.999 	&	--12.495 \\
	$^{230}$Th$\to$$^{206}$Hg+$^{24}$Ne	&	57.76 	&	24.63	&	8 	&	14 	&	0.217	&	--5.661 	&	--11.349 \\
	$^{231}$Pa$\to$$^{207}$Tl+$^{24}$Ne 	&	60.41 	&	22.89 	&	9 	&	14 	&	0.214 	&	--6.271	&	--12.074 \\
	$^{232}$U$\to$ $^{208}$Pb+$^{24}$Ne &62.31 	&	20.39	&	10 	&	14 	&	0.207 	&	--5.924 	&	--11.195 
	\\
	$^{233}$U$\to$ $^{209}$Pb+$^{24}$Ne	&60.49 
	&	24.84	&	10 	&	15 	&	0.210 	&	--6.672 		&	--13.145
	\\
	$^{234}$U$\to$$^{210}$Pb+$^{24}$Ne  &58.82 
	&	25.93 	&	10 	&	16 	&	0.214 	&	--6.313 		&	--11.904
	\\
	$^{233}$U$\to$ $^{208}$Pb+$^{25}$Ne	&60.70 
	&	24.84	&	10 	&	15 	&	0.210 	&	--6.962 	&	--13.126
	\\
	$^{234}$U$\to$$^{208}$Pb+$^{26}$Ne  &59.41 
	&	25.93	&	10 	&	16 	&	0.214 	&	--6.862	& --12.074
	\\
	$^{234}$U$\to$$^{206}$Hg+$^{28}$Mg 
	&74.11 	&	25.53	&	10 	&	16 	&	0.214 	&	--7.411 	&	--13.928 
	\\
	$^{236}$Pu$\to$$^{208}$Pb+$^{28}$Mg	&	79.67 	&	21.52	&	12 	&	16	&	0.203 	&	--7.689 	&	--14.132
	\\
	$^{238}$Pu$\to$$^{210}$Pb+$^{28}$Mg  &	75.91 	&	25.7 	&	12 	&	18 	&	0.210 	&	--7.547 	&	--13.903
	\\
	$^{238}$Pu$\to$$^{208}$Pb+$^{30}$Mg	&	76.79 	&	25.7 &	12 	&	18 	&	0.210 	&	--9.452 	&	--14.561 
	\\
	$^{238}$Pu$\to$$^{206}$Hg+$^{32}$Si	&	91.19 	&	25.28 &		12 	&	18	&	0.210	&	--11.039 	&	--15.625
	\\
	$^{242}$Cm $\to$$^{208}$Pb+$^{34}$Si   &	96.54 	&	23.15	 &	14 	&	20 	&	0.207 	&	--11.366 	&	--16.511 	\\	
\end{longtable*}
\endgroup

\begin{figure}[h]\centering
	\includegraphics[width=8.6cm]{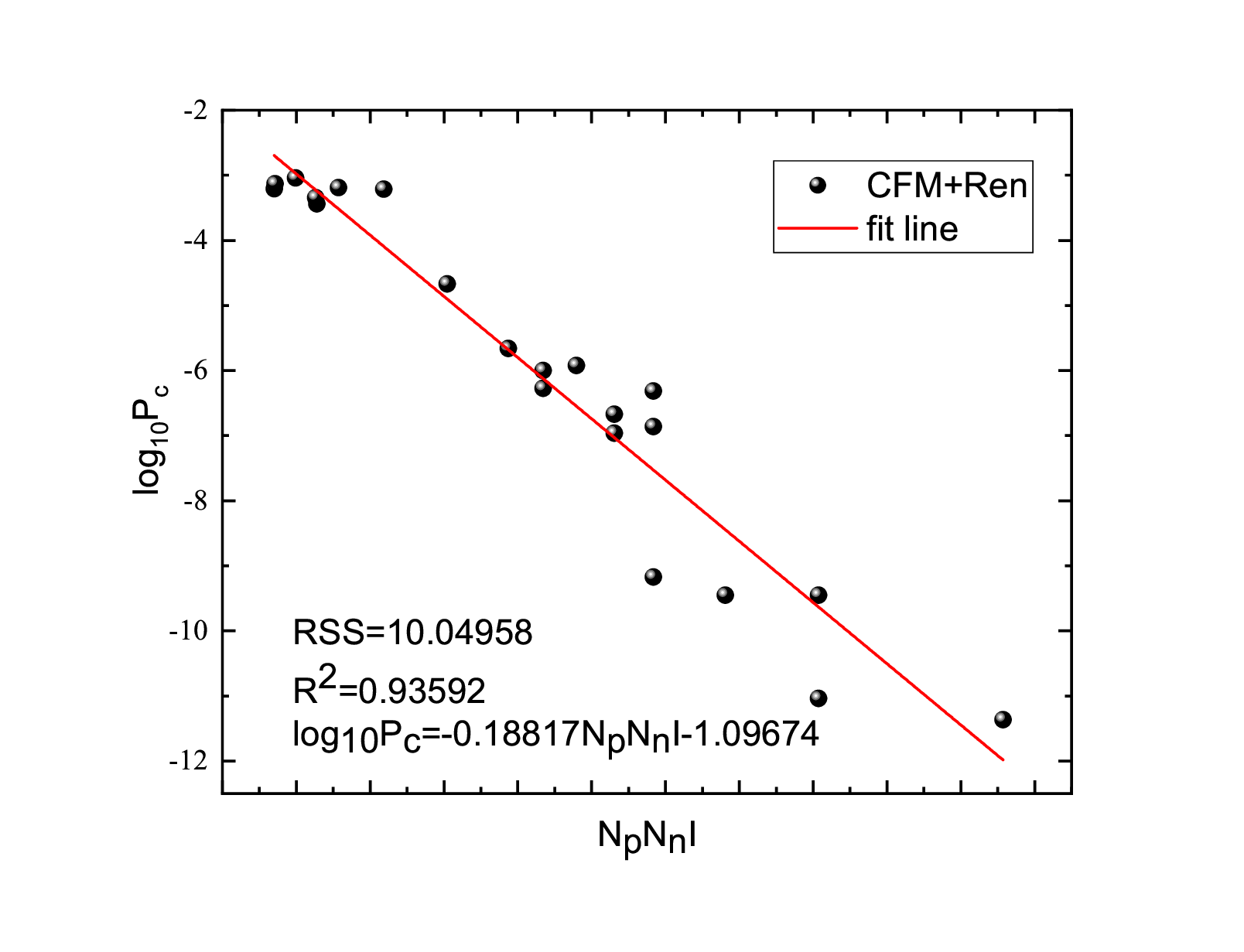}
	\caption{(color online) The linear relationship of the cluster radioactivity preformation probability obtained by microscopic approach in logarithmic form versus the quantity $NpNnI$.}
	\label{fig 3}
\end{figure}
\begin{figure}[h]\centering
	\includegraphics[width=8.6cm]{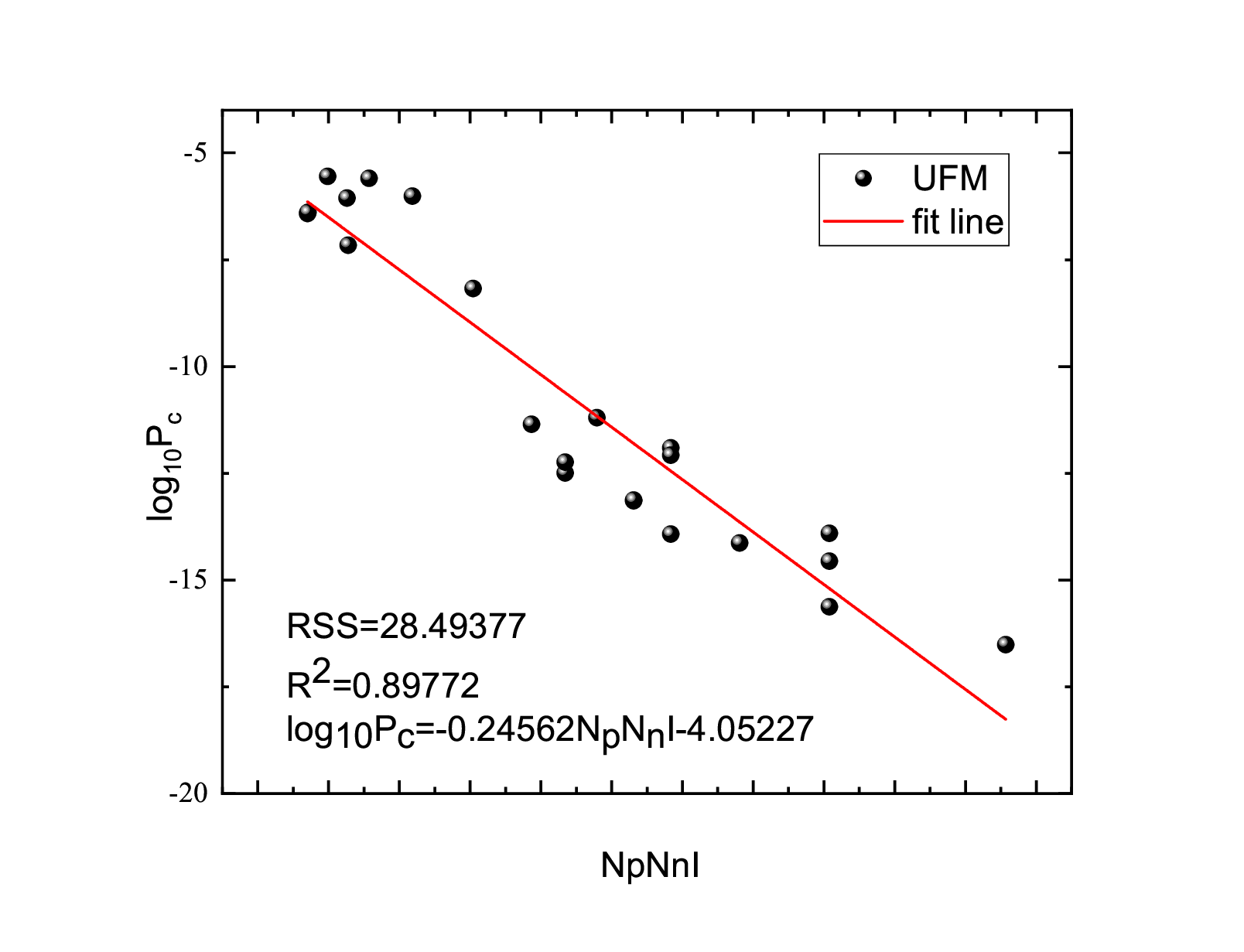}
	\caption{(color online) The linear relationship of the cluster radioactivity preformation probability obtained by model-dependent approach in logarithmic form versus the quantity $NpNnI$.}
	\label{fig 4}
\end{figure}

\textcolor{blue}{}
\begin{equation}
{\rm log_{10}} P_{c}=a\frac{N_{p}N_{n}}{N_{0}+Z_{0}}+b,
\end{equation}
where $a$ and $b$ are the adjustable parameters extracted from the fittings of Fig.\ref{fig 1} and Fig.\ref{fig 2}. 

Moreover, we introduce the statistical quantities, the minimum residual sum of squares (RSS) which represents the sum of squares due to error as well as the coefficient of determination $R^{2}$, to estimate the degree of fitting. In the present work, RSS can be defined as
\begin{eqnarray}\label{16}
		{\rm RSS} = \sum{({\rm {log}}_{10}{T_{1/2}^{\rm{calc}}}-\hat{{\rm {log}} _{10}{T_{1/2}^{\rm{calc}}}})^2},
\end{eqnarray}
 \setlength{\parindent}{0pt} where ${\rm {log}}_{10}{T_{1/2}^{\rm{calc}}}$ and $\hat{{\rm {log}}_{10}{T_{1/2}^{\rm{calc}}}}$ are the logarithmic form of calculated cluster radioactivity half-lives and the value for the corresponding spot on the regression straight line. The total sum of squares (TSS) can be obtained by
\begin{eqnarray}\label{16}
		{\rm TSS} = \sum{({\rm {log}}_{10}{T_{1/2}^{\rm{calc}}}-\bar{{\rm {log}}_{10}{T_{1/2}^{\rm{calc}}}})^2},
\end{eqnarray}
 \setlength{\parindent}{0pt} where $\bar{{\rm {log}}_{10}{T_{1/2}^{\rm{calc}}}}$ denotes the average value for the logarithmic form of calculated cluster radioactivity half-lives. To this end, the coefficient of determination $R^{2}$ can be expressed as

\begin{eqnarray}\label{16}
		 R^{2} =\rm \frac{TSS-RSS}{TSS}= 1-\frac{RSS}{TSS}.
\end{eqnarray}
The smaller the value of RSS and the larger the value of $R^{2}$ approaching 1, the better the degree of fit. $a$ and $b$ are fitted with the minimum RSS and $R^{2}$ $\approx$ 1. The detailed corresponding values of $a$, $b$, RSS and $R^{2}$ are given in Fig.\ref{fig 1} and Fig.\ref{fig 2}.

The general behavior of $P_{c}$ in logarithmic form shows a decrease with the increase of $N_{p}N_{n}$. Furthermore, for the value of parent nucleus isospin asymmetry depends on the doubly magic core at ($N_{0}$, $Z_{0}$), we size up the $P_{c}$ by incorporating the isospin asymmetry into $N_{p}N_{n}$, which can be expressed as 
\begin{equation}
{\rm log}_{10}P_{c}=cN_{p}N_{n}I+d,
\end{equation}
where $c$ and $d$ are the adjustable parameters by fitting to Fig.\ref{fig 3} and Fig.\ref{fig 4} whose RSS value is minimum and coefficient of determination $R^{2}$ $\approx$ 1. The plots of the dependence for the $P_{c}$ in logarithmic form with $N_{p}N_{n}I$ have been presented in Fig.\ref{fig 3} and Fig.\ref{fig 4}. From these two figures, the linear correlations of the deduced cluster radioactivity preformation probability in logarithmic form with the valence nucleons (holes) numbers become more distinct when they are plotted as a function of $N_{p}N_{n}I$. It is clearly to see that cluster radioactivity preformation probability in logarithmic form are proportional to the products of effective numbers of valence particles (holes) $N_{p}N_{n}$ as well as the multiplication of the isospin asymmetry with this product $N_{p}N_{n}I$.
 
 \begingroup
 \renewcommand*{\arraystretch}{1.3}
 \setlength{\tabcolsep}{0.00005mm}
 \begin{longtable*}{ccccccc}	
 	\caption{Cluster radioactivity preformation probability in different formulas and models. See text for details.}
 	\label{Tab2} \\
 	\hline 
 	\hline 
 	{Decay}&\qquad$Q_{c}$(MeV)\qquad&\qquad\rm$T^{exp}_{1/2}$\qquad&\qquad CFM\qquad&\qquad {UFM}\qquad&\qquad WZM\qquad&\qquad WZF\qquad\\
 	\hline
 	\endfirsthead
 	\multicolumn{7}{c}%
 	{{\tablename\ \thetable{} -- continued from previous page}} \\
 	\hline
 	\hline 
 	{Decay}&\qquad$Q_{c}$(MeV)\qquad&\qquad\rm $T^{exp}_{1/2}$\qquad&\qquad CFM\qquad&\qquad {UFM}\qquad&\qquad WZM\qquad&\qquad WZF\qquad\\
 	\hline
 	\endhead
 	\hline \multicolumn{7}{r}{{Continued on next page}} \\
 	\endfoot
 	\hline \hline
 	\endlastfoot
  	$^{221}$Fr$\to$ $^{207}$Tl+$^{14}$C	&	31.29	&14.56 & --3.204
 	&--6.420&--10.00 &--10.18
 	
 	\\
 	$^{221}$Ra$\to$$^{207}$Pb+$^{14}$C	& 32.40 	&$	13.39$ 	&--3.126			&--6.388	&--9.95 &--10.18	\\
 	$^{222}$Ra$\to$$^{208}$Pb+$^{14}$C & 33.05	&11.22 &--3.037	&--5.547	&--8.98 &--10.94 
 	\\
 $^{223}$Ra$\to$$^{209}$Pb+$^{14}$C &31.83	&15.05		&--3.440	&--7.160	&--10.68 &--11.68
 	\\
 	
 	$^{224}$Ra$\to$$^{210}$Pb+$^{14}$C&	30.53 &15.87		 &--3.186	&--5.593	 &--9.06&--12.39	 \\
 	$^{226}$Ra$\to$ $^{212}$Pb+$^{14}$C & 28.20 	&21.20	 	&--3.211	&--6.014&--9.52&--13.75
 	\\
 $^{223}$Ac$\to$$^{209}$Bi+$^{14}$C	& 33.06	& 12.60 &--3.341&--6.060 &--9.55 &--11.68
 	\\
 $^{228}$Th$\to$$^{208}$Pb+$^{20}$O	&44.72	&20.73 &--4.670	&--8.174	 &--12.76&--15.02
 	\\
 $^{231}$Pa$\to$$^{208}$Pb+$^{23}$F	& 51.88	&26.02 &	--5.999	&--12.495	 &--17.60&--16.76
 	\\$^{230}$Th$\to$$^{206}$Hg+$^{24}$Ne	&57.76	&24.63		&--5.661	&--11.349	 &--16.62&--16.20
 	\\
 	$^{231}$Pa$\to$$^{207}$Tl+$^{24}$Ne &60.41	&22.89 	&--6.271		&--12.074 &--17.50 &--16.76
 	\\
 	$^{232}$U$\to$ $^{208}$Pb+$^{24}$Ne & 62.31	&20.39	&--5.924&--11.195 &--16.43	 
 	&--17.30
 	\\
 	$^{233}$U$\to$ $^{209}$Pb+$^{24}$Ne	&60.49
 	&24.84	&	--6.672	&--13.145	&--18.44 &--17.82
 	\\
 	$^{234}$U$\to$$^{210}$Pb+$^{24}$Ne &58.82
 	&25.93&--6.313		&--11.904&--17.18 &--18.32
 	\\$^{233}$U$\to$ $^{208}$Pb+$^{25}$Ne	&60.70
 	&24.84&	--6.962	&--13.126	&--18.56&--17.82
 	\\$^{234}$U$\to$$^{208}$Pb+$^{26}$Ne &59.41
 	&25.93	&--6.862		& --12.074&--17.65	&--18.32
 	\\
            $^{234}$U$\to$$^{206}$Hg+$^{28}$Mg
 	&74.11	&25.53		&--7.411&--13.928	&--19.82 &--18.32
 	\\
 $^{236}$Pu$\to$$^{208}$Pb+$^{28}$Mg	& 79.67	& 21.52	&--7.689	&--14.132	&--19.98&--19.26
 	\\
  $^{238}$Pu$\to$$^{210}$Pb+$^{28}$Mg &75.91	&25.70	 &--7.547		&--13.903 &--19.80 &--20.13
 	\\
 	$^{238}$Pu$\to$$^{208}$Pb+$^{30}$Mg	&	76.79	&25.70	 	&--9.452		&--14.561	 &--20.72&--20.13
 	\\$^{238}$Pu$\to$$^{206}$Hg+$^{32}$Si	&91.19	&25.28	 	&--11.039		&--15.625	&--22.09 &--20.13
 	\\$^{242}$Cm $\to$$^{208}$Pb+$^{34}$Si   &96.54	&23.15	&--11.366		&--16.511 &--23.20 &--21.64
 	\\	
 \end{longtable*}
 \endgroup
 
For further investigating the preformation mechanics for cluster radioactivity, we also compare the $P_{c}$ results extracted from the experimental decay energy and half-life as well as the results obtained by using the parent-mass-number dependence analytical formula and model proposed by K. Wei and H. F. Zhang \cite{Phys. Rev. C 96 021601} with our work. The calculated results are  well listed in Table. \ref{Tab2}. In this table, the first three columns are the same as Table. \ref{Tab1}. The fourth to seventh columns are shown the cluster radioactivity preformation probability in logarithmic form obtained by using CFM, UFM, Wei's model as well as Wei's formula denoted as CFM, UFM, WZM and WZF, respectively.  \begin{figure}[h]\centering
	\includegraphics[width=8.6cm]{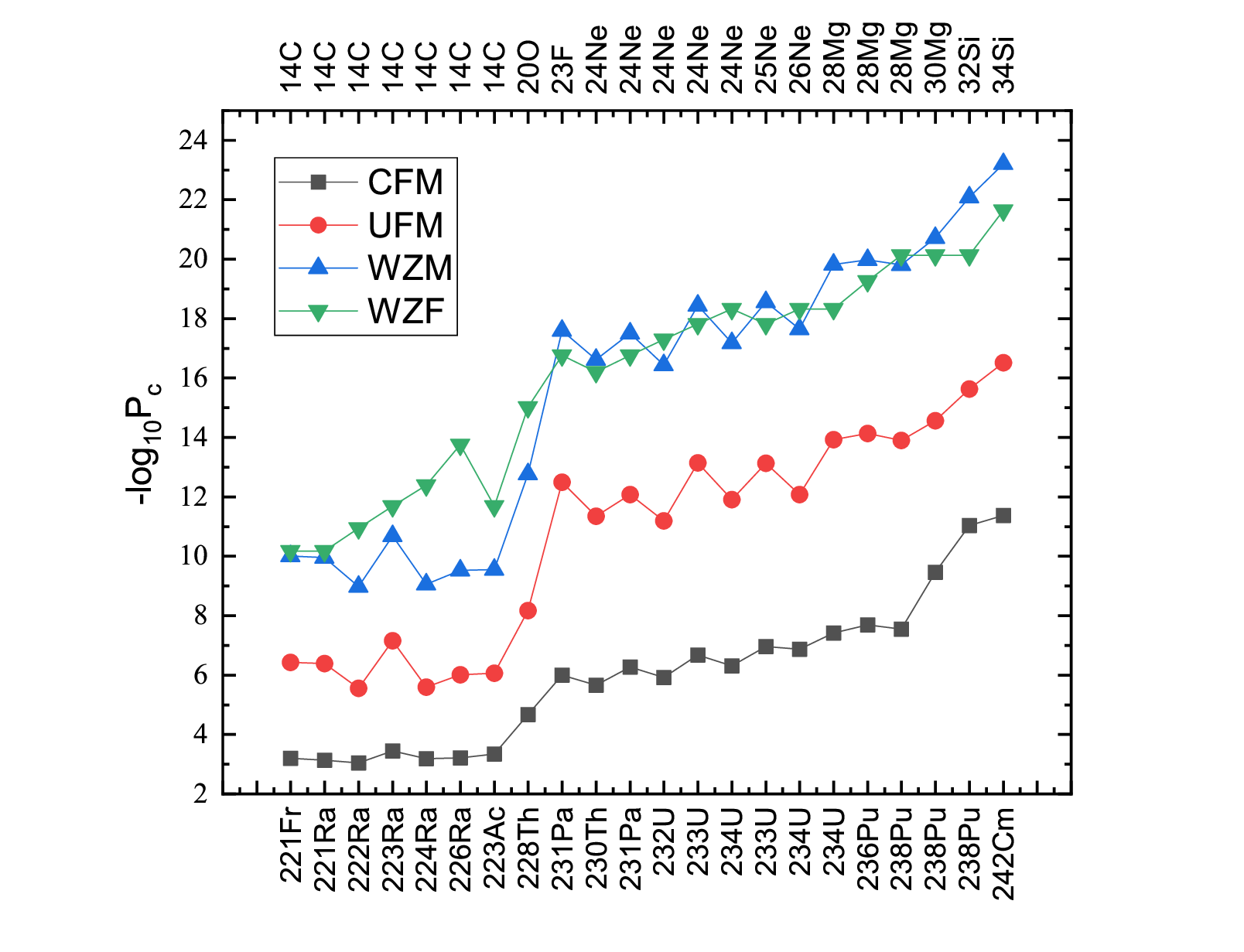}
	\caption{(color online) The tendency of the negative values for cluster radioactivity preformation probability obtained by using different models and formulas extracted from Table. \ref{Tab2} in logarithmic form.}
	\label{fig 5}
\end{figure}It is obviously to find that the general tendencies of the preformation probability in logarithmic form are similar to the pattern for the variations of $P_{c}$ observed from Table. \ref{Tab1}. For a more deeper insight into this phenomenon, the tendency of the variations for $P_{c}$ in logarithmic form obtained by using CFM, UFM, Wei's model and Wei's formula has been plotted in Fig.\ref{fig 5}. From this figure, it is easily to see that the values of $P_{c}$ are generally decreasing with the increasing of the mass number parent nuclei as well as the emitted cluster. For neutrons pairing is more influential than protons in cluster radioactivity and most of the cluster emitters as well as the emitted cluster are neutron-rich nuclides, it is less possibly for the cluster to be formed in the parent nuclei in trans-lead region far away from neutron shell closure at 126 \cite{Phys. Scr. 95 075303}.

  The results of our work further validate that cluster radioactivity is closely to the nuclear shell effect. Investigation for the cluster radioactivity preformation probability explored in the scheme of $N_{p}N_{n}$ could well simply reflect the shell effects of nuclear structure and can be easily acquired as they are parameterized in the simplified functions of $N_{p}$ and $N_{n}$. We hope this study could be useful for studying cluster radioactivity and further probing the nuclear structure in the trans-lead region.
\section{Summary}
\label{section 4}

In summary, we systematically investigate the dependence of cluster radioactivity preformation probability versus valence protons and neutrons both microscopically and model-dependently. In the microscopic approach, based on our previous work, the cluster radioactivity preformation probability is dealt with cluster formation model (CFM) combined with the exponential relationship of $P_{c}$ to the $\alpha$ decay preformation probability $P_{\alpha}$ when the number of the emitted cluster $A_{c}$ $\leq$ 28. Whereas $A_{c}$ $\ge$ 28, $P_{c}$ is obtained through the charge-number dependence of $P_{c}$ on the decay products proposed by Ren $et$ $al.$. In the model-dependent approach, $P_{c}$ is extracted through the ratios from the calculated cluster radioactivity half-lives to experimental ones, while the cluster radioactivity half-lives are obtained within the framework of UFM. Both of the results have shown the cluster radioactivity preformation probability $P_{c}$ in the logarithmic form is proportional to the products of the valence protons and neutrons as well as the multiplication of this product with the isospin asymmetry. We also compare the results obtained by using the parent-mass-number dependence analytical formula as well as the model proposed by K. Wei and H. F. Zhang with our work and make a discussion about preformation mechanic for cluster radioactivity.

\begin{acknowledgments}
 We thank K. Wei and Dr. J. G. Deng for their support and helpful discussions. This work is supported in part by the National Natural Science Foundation of China (Grant No.12175100 and No.11975132), the construct program of the key discipline in hunan province, the Research Foundation of Education Bureau of Hunan Province, China (Grant No.18A237), the Shandong Province Natural Science Foundation, China (Grant No.ZR2022JQ04), the Opening Project of Cooperative Innovation Center for Nuclear Fuel Cycle Technology and Equipment, University of South China (Grant No.2019KFZ10), the Innovation Group of Nuclear and Particle Physics in USC, Hunan Provincial Innovation Foundation for Postgraduate (Grant No.CX20210942).
\end{acknowledgments}


%

\end{document}